\documentclass[conference]{IEEEtran}
\IEEEoverridecommandlockouts
\usepackage{amsmath,amssymb,amsfonts}
\usepackage{algorithmic}
\usepackage{graphicx}
\usepackage{subcaption}
\usepackage{textcomp}
\usepackage{xcolor}
\usepackage{cite}
\def\BibTeX{{\rm B\kern-.05em{\sc i\kern-.025em b}\kern-.08em
    T\kern-.1667em\lower.7ex\hbox{E}\kern-.125emX}}
\begin{document}

\title{Analyzing Strategies for Dynamical Decoupling Insertion on IBM Quantum Computer
\thanks{This work is funded by the QuantUM Initiative of the Region Occitanie, University of Montpellier and IBM Montpellier.}
}

\author{\IEEEauthorblockN{Siyuan Niu}
\IEEEauthorblockA{\textit{LIRMM, University of Montpellier} \\
161 Rue Ada, Montpellier, France \\
siyuan.niu@lirmm.fr}
\and
\IEEEauthorblockN{Aida Todri-Sanial}
\IEEEauthorblockA{\textit{LIRMM, University of Montpellier, CNRS} \\
161 Rue Ada, Montpellier, France \\
aida.todri@lirmm.fr}

}

\maketitle

\begin{abstract}
Near-term quantum devices are subject to errors and decoherence error is one of the non-negligible sources. Dynamical decoupling (DD) is a well-known technique to protect idle qubits from decoherence error. However, the optimal approach to inserting DD sequences still remains unclear. In this paper, we identify different conditions that lead to idle qubits and evaluate strategies for DD insertion under these specific conditions. Specifically, we divide the idle qubit into crosstalk-idle or idle-idle qubit depending on its coupling with other qubits and report the DD insertion strategies for the two types of idle qubits. We also perform Ramsey experiment to understand the reasons behind the strategy choice. Finally, we provide design guidelines for DD insertion for small circuits and insights for large-scale circuit design. 
\end{abstract}


\section{Introduction}
Quantum computing has become an active research field in recent years. It is expected to address some classically intractable computational problems, such as integer factorization~\cite{shor1999polynomial}, optimization~\cite{farhi2014quantum}, quantum simulation~\cite{georgescu2014quantum}, etc. Quantum supremacy has been achieved for superconducting and photonic device~\cite{arute2019quantum, zhong2020quantum}. IBM released the largest universal
quantum machine with 127 superconducting qubits. However, the near-term quantum hardware is still regarded as Noisy Intermediate-Scaled Quantum (NISQ) machine~\cite{preskill2018quantum} with hardware limitations depending on the quantum technology used. Qubits have unavoidable error rates, which makes finding ``real'' quantum advantage difficult. It is predicted to require more than one thousand physical qubits to encode an error-free logical qubit for error correction~\cite{fowler2012surface}, which is beyond the current quantum technology. Instead, quantum error mitigation was proposed to reduce the noise impact and improve the fidelity of NISQ applications~\cite{endo2018practical}.

There are different error sources on NISQ devices, such as gate error, readout error, decoherence error and crosstalk. In our paper, we focus on the decoherence error which occurs on the idle qubits, where errors accumulate over time even if there is no operation, also called idling error. In superconducting quantum devices, quantum operations are performed in parallel if applied to different qubits. But these parallel operations lead
to a high probability of idle qubits introducing decoherence errors, mainly for two main reasons: (1) Gate latencies vary across different qubits and gate types. For example, for IBM Q 27 Montreal, the average gate duration for two-qubit gates is 2.6x larger than  for single-qubit gates and the longest two-qubit gate duration is 2.3x larger than the shortest one. (2) The nearest-neighbor connectivity of superconducting device needs \texttt{SWAP} gate to realize non-adjacent connections, which is composed of three \texttt{CNOTs} and requires long duration to execute. 
%

Dynamical Decoupling (DD) is one of the simplest methods for error mitigation without any extra circuit overhead, and is dedicated to the decoherence error suppression~\cite{souza2012robust}. The basic idea of DD is to repetitively insert DD sequences to the idle qubit and make sure to return back to the qubit's original state after insertion. There are various DD sequences, such as CPMG~\cite{meiboom1958modified}, XY4~\cite{viola1999dynamical}, KDD~\cite{souza2011robust}, etc. It has been demonstrated to improve circuit fidelity on IBM and Rigetti platforms using XY4 decoupling sequence~\cite{pokharel2018demonstration}. Also, DD has been proved to have the capability of suppressing ZZ-crosstalk in superconducting qubits~\cite{tripathi2021suppression}. However, DD insertion is not trivial such that it cannot always improve application fidelity. Whether one can obtain favorable DD insertion depends on many factors such as the given quantum application, the choice of DD strategy, the number of repetitions of DD sequences, etc. Several quantum software tools were developed to help design DD insertion strategies to obtain better circuit fidelity. Das et al. proposed ADAPT~\cite{das2021adapt}, an adaptive DD framework to insert DD pulses to a subset of idle qubits. Moreover, Ravi et al. introduced VAQEM~\cite{ravi2021vaqem}, which dynamically selects the length of DD sequence for variational quantum algorithms. These two techniques need a large overhead of tuning DD sequences with extra circuits. Recently, Niu et al. reported the impact of different DD strategies on well-known quantum applications to provide insights for application-specific error mitigation using DD technique~\cite{niu2022pulse}. However, these previous methods only concentrate on application-level DD insertion instead of understanding and analyzing the conditions that lead to idle qubits, hence decoherence errors.

In order to find the optimal DD insertion strategy, one needs first to understand when the idle qubits usually occur in quantum circuits. In our paper, we address this issue by investigating the conditions that cause idle qubits and evaluating the strategies for DD insertion under specific configurations to improve the circuit fidelity. We create a three-qubit circuit to discuss the DD insertion strategy and perform the experiments on IBM Q 7 Lagos. Our major contributions can be listed as follows:

\begin{itemize}
	\item We identify several conditions leading to idle qubits: \texttt{CNOT}-induced idle, \texttt{SWAP}-induced idle (multiple \texttt{CNOTs}), and combine them with additional delays that might be caused by other operations.
	\item For each condition, we assess different DD insertion strategies and report their impact on idle-idle qubit and crosstalk-idle qubit.
	\item We perform the Ramsey experiment to explain the results and provide guidelines for DD insertion technique on small-scale quantum circuits as well as insights for large-scale circuits. 
	
\end{itemize}

\section{Background}
\subsection{NISQ hardware} 
Today's quantum device is in the NISQ era with hardware limitations and erroneous operations. For superconducting quantum devices, one of the currently leading quantum technologies and the focus of this paper, the connectivity can only be realized between neighbor qubits. A quantum circuit needs to be transpiled to satisfy the nearest-neighbor connectivity by inserting \texttt{SWAP} gates before executing on the real hardware, known as the qubit mapping problem~\cite{li2019tackling,niu2020hardware}. Moreover, the near-term quantum machine is not error corrected and we broadly classify the error sources as follows: (1) Operational error. It includes gate error and SPAM error. (2) Decoherence error. A qubit can only maintain its state for a limited amount of time due to its fragile nature. (3) Crosstalk. The state of a qubit might be corrupted by the simultaneous operations occurring on its neighbor qubits. All of these errors prevent a quantum circuit to obtain reliable results. Several quantum error mitigation techniques have been proposed, such as zero noise extrapolation~\cite{temme2017error}, Pauli twirling~\cite{cai2019constructing}, measurement error mitigation~\cite{bravyi2021mitigating}, dynamical decoupling~\cite{souza2012robust}, etc.

\subsection{Dynamical decoupling}
DD is widely used to suppress the impact of decoherence errors caused by the interaction between the system and the environment. Instead of keeping a qubit in the idle state, DD continuously inserts pulses during the idle time and makes the qubit return back to its original state after DD insertion such that the overall DD sequences act like an identity gate. DD has shown benefits in improving circuit fidelity on IBM, Rigetti, and Google quantum devices~\cite{pokharel2018demonstration,chen2021exponential}. 

Different DD strategies have been proposed, such as CPMG~\cite{meiboom1958modified}, XY4~\cite{viola1999dynamical}, UDD~\cite{uhrig2007keeping}, KDD~\cite{souza2011robust}, etc. In our paper, we evaluate one of the most frequently used DD strategies: CPMG. It was introduced to reduce the inhomogeneous effects of the environment. The basis DD cycle to constitute the CPMG sequence is $X-X$.

\subsection{Motivational example}
\label{sec:motivation}

The idle qubit can be divided into two types: (1) idle-idle qubit, where no operation is applied to its neighbor qubits in parallel. (2) crosstalk-idle qubit, where simultaneous operations are occurring on its neighbor qubits such that the target qubit has a probability of being influenced by crosstalk. It has been demonstrated that two-qubit operation is the dominant crosstalk source for simulataneous operations~\cite{murali2020software}. Therefore, we ignore the case when crosstalk comes from parallel single-qubit operations and only consider the crosstalk due to \texttt{CNOT} gate.

\begin{figure}[h]
	\centering
	
	\includegraphics[scale=0.6]{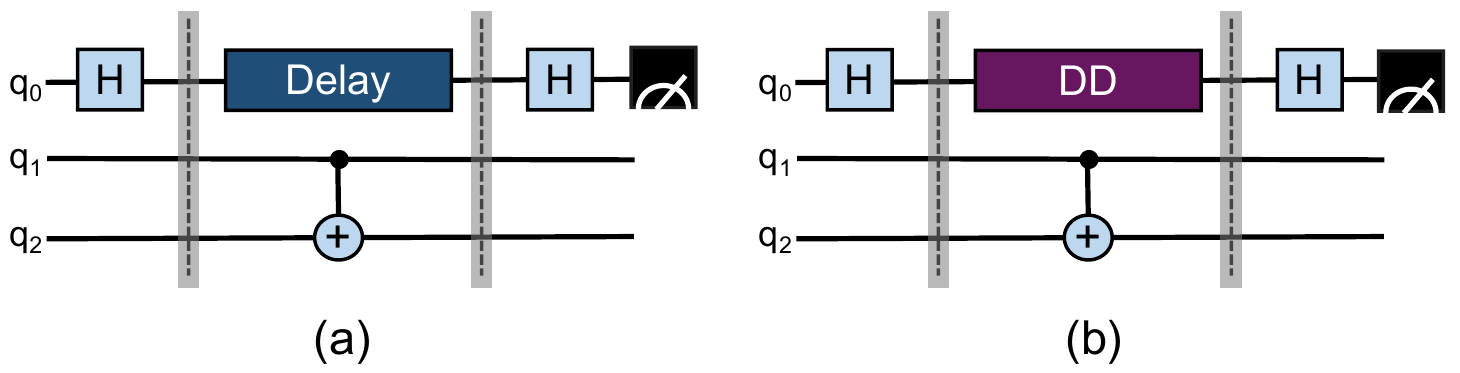}
	\label{fig:multi}
	
	\caption{A motivational example to show the DD impact on circuit fidelity where the main qubit $q_0$ is under (a) free evolution and (b) DD-inserted evolution.
	}
	\label{fig:motivation}
	
\end{figure}

\begin{figure}
	\centering
	\begin{subfigure}{0.45\columnwidth}
		\centering
		\includegraphics{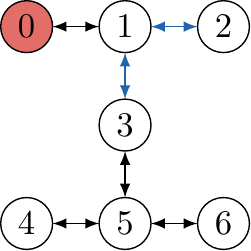}
		\label{fig:idle1}
		\caption{}
	\end{subfigure}
	\hfil
	\begin{subfigure}{0.45\columnwidth}
		\centering
		\includegraphics{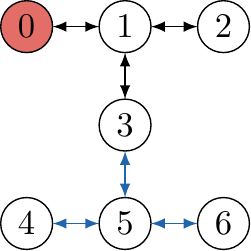}
		\label{fig:idle2}
		\caption{}
	\end{subfigure}
	
	\caption{ The main qubit $q_0$ mapped to $Q_0$ in red color performs as (a) crosstalk-idle qubit, or (b) idle-idle qubit, when the spectator qubits are mapped to one of the qubit pairs whose link is highlighted in blue color in (a) (resp. (b)).
	}
	\label{fig:idles}
	
\end{figure}

Here, we show a motivational example in Fig.~\ref{fig:motivation} to characterize the impact of DD sequence on idle-idle qubit and crosstalk-idle qubit. We create a 3-qubit circuit, where $q_0$ is the main qubit to measure, $q_1$ and $q_2$ are two spectator qubits to apply \texttt{CNOT} gate. First, the main qubit is under a free evolution where the duration of the delay (idle time) is the same as the \texttt{CNOT} gate applied to the spectator qubits (Fig.~\ref{fig:motivation}(a)). Second, the main qubit is under a DD-inserted evolution during the idle time (Fig.~\ref{fig:motivation}(b)). Two Hadamard gates are inserted at the beginning and the end of the circuit, respectively. The first $H$ gate is used to put the main qubit in a superposition state. The second one is to undo the initial $H$ gate such that the ideal state of the main qubit should be $|0 \rangle$ if no errors occur during the idle time. The middle part inside the barriers is repeated from 10 to 50 times with a step of 10 to show the qubit state evolution across time. In our case, we evaluate the performance of CPMG so that a $X-X$ sequence is inserted and only repeated once during the overall idle time.
Note that, the number of DD sequences inserted during the idle time can influence the circuit fidelity~\cite{ravi2021vaqem}, but this is not in the scope of this paper. 
We map the main qubit of the circuit to each physical qubit on the quantum hardware and the spectator qubits to two physically connected qubits. Note that, we always choose the hardware-native \texttt{CNOT} so that no additional single-qubit gate is required when executed on real quantum hardware. For example, if we map the main qubit $q_0$ to $Q_0$ in IBM Q 7 Lagos shown in Fig.~\ref{fig:idles}, there are five possibilities for spectator qubits: $\{Q_1, Q_2\}$,$\{Q_1, Q_3\}$,$\{Q_3, Q_5\}$,$\{Q_4, Q_5\}$,$\{Q_5, Q_6\}$. Based on the impact of spectator qubits on the main qubit, the main qubit can be categorized into two types of idle qubit: 
 (1) crosstalk-idle qubit, where one of the spectator qubit is a neighbor qubit to the target main qubit and the spectator qubits is one pair from $\{Q_1, Q_2\}$,$\{Q_1, Q_3\}$. (2) idle-idle qubit, where the spectator qubits are further away from the target main qubit (the distance between spectator qubits and main qubit is more than one on hardware topology), and they can be from $\{Q_3, Q_5\}$,$\{Q_4, Q_5\}$,$\{Q_5, Q_6\}$. 

\begin{figure}
	\centering
	\begin{subfigure}{0.45\columnwidth}
		\centering
		\includegraphics[scale=0.3]{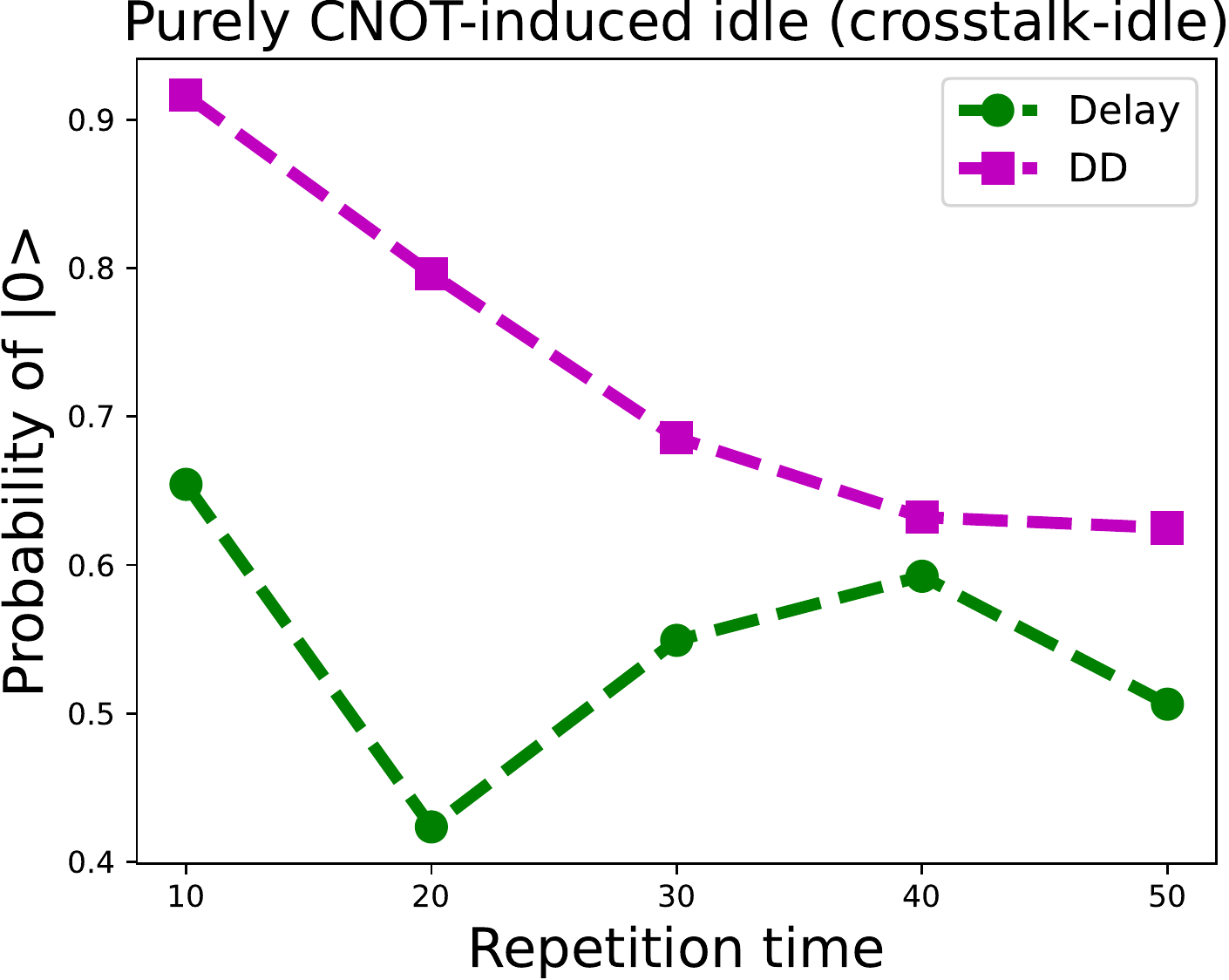}
		\label{fig:DD-cnot-crosstalk}
		\caption{}
	\end{subfigure}
	\hfil
	\begin{subfigure}{0.45\columnwidth}
		\centering
		\includegraphics[scale=0.3]{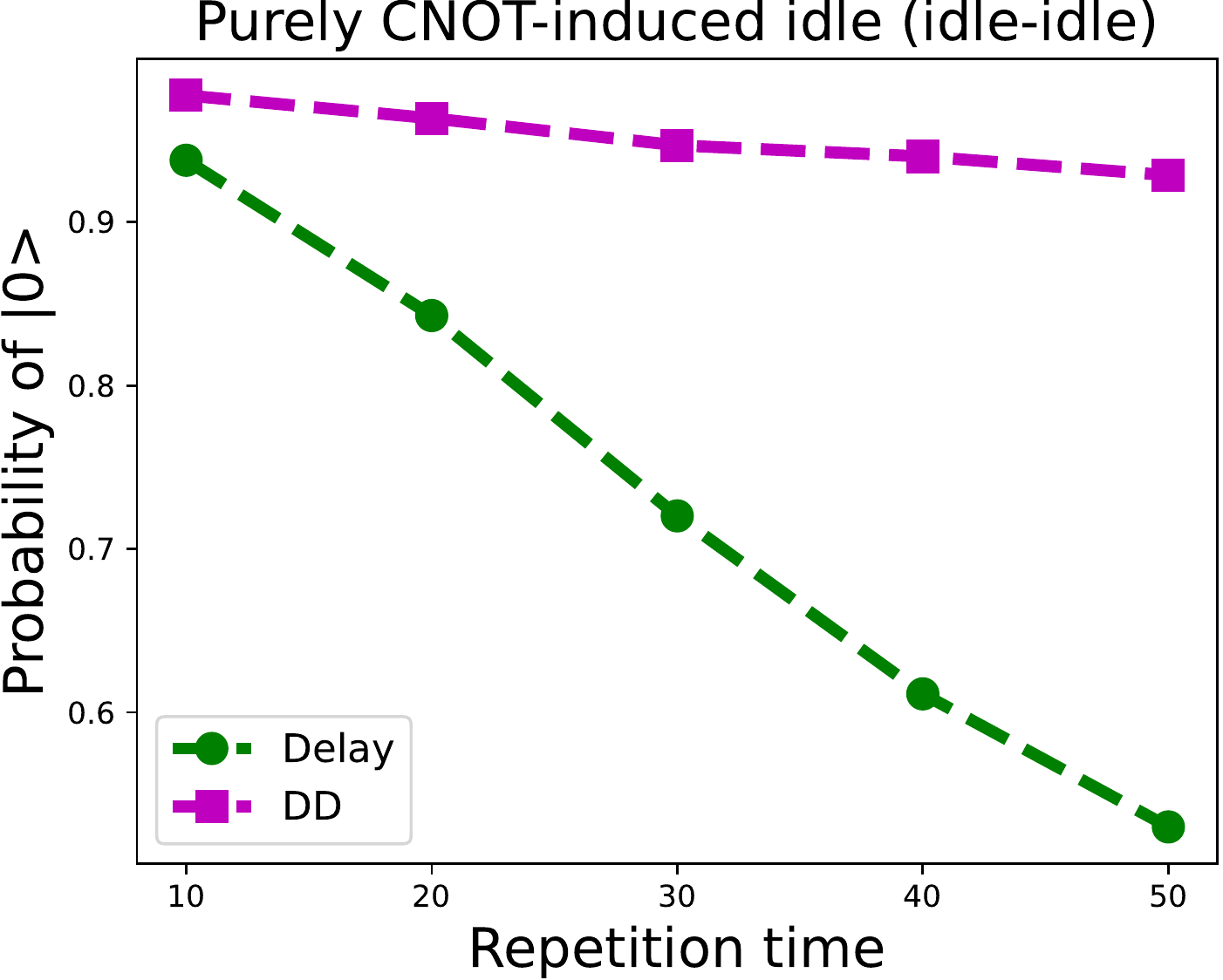}
		\label{fig:DD-cnot-no-crosstalk}
		\caption{}
	\end{subfigure}
	
	\caption{The average probability of obtaining $|0 \rangle$ when the main qubit $q_0$ is considered as (a) crosstalk-idle qubit. (b) idle-idle qubit.
	}
	\label{fig:DD-cnot}
	
\end{figure}

The results of the motivational example are shown in Fig.~\ref{fig:DD-cnot}. The crosstalk impact is significant such that the probability of obtaining $|0 \rangle$ for crosstalk-idle qubit decreases much faster than idle-idle qubit. When the quantum coherence is completely lost, the state of the main qubit becomes $\rho = \frac{1}{2}(|0 \rangle \langle 0 \rangle + |1 \rangle \langle 1 \rangle)$ so that the probability of obtaining $|0 \rangle$ is approximately 0.5 in the end. 
DD is favorable to increase circuit fidelity for the two types of idle qubits. 
 

\section{Related works}
Even though DD has shown to play an important role in reducing decoherence error on different quantum machines, there is still no clear answer to the optimal way to insert DD pulses. Here are some interesting questions: (1) Which DD strategy performs the best for a specific platform or application? (2) Should we insert DD sequences to all the idle qubits? (3) What is the optimal number of the DD pulses during the idle time? (4) What conditions lead to idle qubits and what is the strategy to insert DD to idle qubits under different conditions? The first question was discussed in~\cite{niu2022pulse}, where the authors comprehensively evaluated the performance of all kinds of DD sequences on well-known quantum applications for IBM quantum devices. Das et al. proposed Adaptive Dynamical Decoupling (ADAPT) to address the second problem~\cite{das2021adapt}. They demonstrated that DD performs better when applied to a subset of qubits rather than all. However, this method requires executing a decoy circuit (a circuit that is structurally similar to the given circuit) with DD insertions on different subsets of qubits to find the best way to insert DD for the input circuit, which introduces a large overhead. Moreover, Ravi et al. introduced VAQEM to solve the third problem~\cite{ravi2021vaqem}, an error mitigation technique that tunes the number of DD sequences inserted in the idle time and selects the best one that boosts the circuit fidelity, which also requires additional circuit measurement overhead. In our paper, we focus on the fourth problem. We determine different conditions that lead to idle qubits and evaluate
 various strategies of inserting DD pulses to find the best solution that benefits the circuit fidelity.  

\begin{figure}
	\centering
	\begin{subfigure}{0.45\columnwidth}
		\centering
		\includegraphics[scale=0.6]{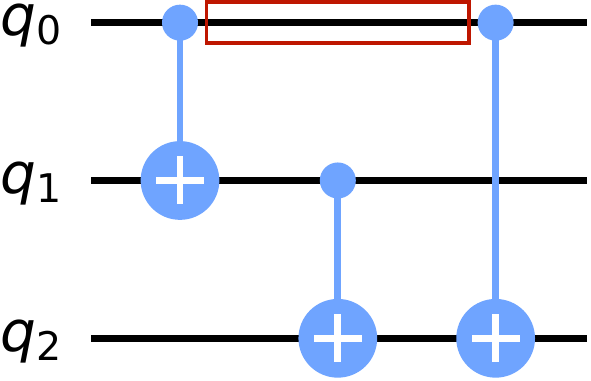}
		\label{fig:cnot-induced-idle}
		\caption{}
	\end{subfigure}
	\hfil
	\begin{subfigure}{0.45\columnwidth}
		\centering
		\includegraphics[scale=0.6]{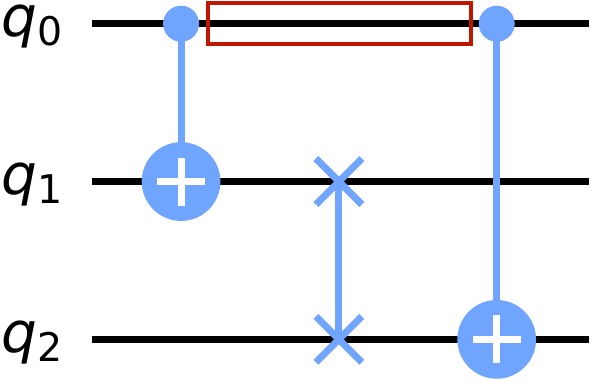}
		\label{fig:swap-induced-idle}
		\caption{}
	\end{subfigure}
	
	\caption{(a) \texttt{CNOT}-induced idle qubit. (b) \texttt{SWAP}-induced idle qubit. The idle time is highlighted in the red frame.
	}
	\label{fig:induced-idle}
	
\end{figure}

\section{Context-induced idle qubit}
For superconducting quantum devices, quantum operations are able to be executed in parallel if applied to different qubits. Parallel scheduling is also preferable since it can reduce the circuit depth so that the decoherence error is decreased as well. However, the gate latency varies across two-qubit gates and their mean latency is much longer than one-qubit gates. Moreover, due to the hardware limitation of the nearest-neighbor connectivity for superconducting quantum devices, \texttt{SWAP} gates are generally needed to execute the two-qubit gates applied to qubits that are not physically connected. Therefore, idle qubits occur frequently in a quantum circuit. In general, there are two conditions leading to idle qubits: (1) \texttt{CNOT}-induced idle qubit caused by \texttt{CNOT} gates executed in parallel, shown in Fig.~\ref{fig:induced-idle}(a). (2) \texttt{SWAP}-induced idle qubit caused by simultaneous \texttt{SWAP} gates, shown in Fig.~\ref{fig:induced-idle}(b). Moreover, the idle time can become longer if some other gates are running in the middle that evolves $q_1$ or $q_2$. 

In this section, we discuss the impact of different DD insertion strategies on these two configurations for two types of idle qubits: idle-idle qubit and crosstalk-idle qubit. We create a 3-qubit circuit similarly to Fig.~\ref{fig:motivation} and discuss the two conditions in the following sections.

\subsection{CNOT-induced Idle Qubit}
\label{sec:cnot-idle}

\begin{figure*}[h]
	\centering
	
	\includegraphics[scale=0.4]{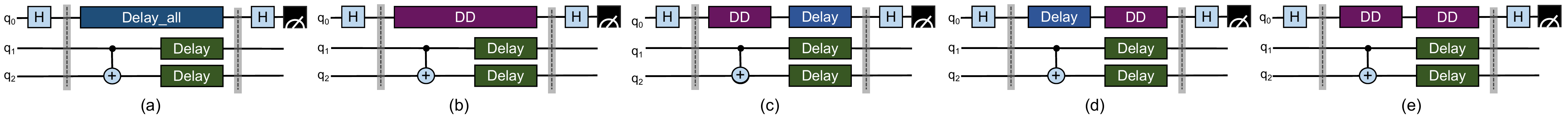}
	
	\caption{Experiments of \texttt{CNOT} and other delay induced idle qubit. (a) Baseline, where $q_0$ is under free evolution during the overall idle time. (b) Inserting DD during the overall idle time. (c) Inserting DD during the \texttt{CNOT} duration and then \emph{Delay}. (d) Inserting \emph{Delay} during \texttt{CNOT} duration and then DD. (d) Inserting double DD separately during \texttt{CNOT} and \emph{Delay}. 
	}
	\label{fig:cnot-delay}
	
\end{figure*}

The basic configuration of purely \texttt{CNOT}-induced idle qubit is shown in section~\ref{sec:motivation}, where inserting DD sequences to idle qubits can significantly improve the circuit fidelity for both idle-idle and crosstalk-idle qubits. Here, we discuss the case when the idle qubit is induced by a \texttt{CNOT} gate as well as some other gates, which are indicated by a \emph{Delay} operation. Overall, there are four DD insertion strategies: (1) Insert DD pulses during the overall idle time (labeled as DD), shown in Fig.~\ref{fig:cnot-delay}(b). (2) First insert DD sequence only during the \texttt{CNOT} gate operation and then a \emph{Delay} operation, which has the same length as the \emph{Delay} in spectator qubits (labeled as DD\_Delay), shown in Fig.~\ref{fig:cnot-delay}(c). (3) First insert \emph{Delay} operation during the parallel \texttt{CNOT} operation and then DD sequence during the \emph{Delay} in spectator qubits (labeled as Delay\_DD), shown in Fig.~\ref{fig:cnot-delay}(d). (4) Insert two independent DD pulses for \texttt{CNOT} and \emph{Delay}, respectively (labeled as DD\_DD), shown in Fig.~\ref{fig:cnot-delay}(e). The length of the \emph{Delay} operation is the same as the \texttt{CNOT} duration to represent the extra idle time induced by other operations. 
We compare them with the baseline (Fig.~\ref{fig:cnot-delay}(a)), where the main qubit $q_0$ is under a free evolution marked as \emph{Delay\_all}, which is the overall duration of the spectator qubits. We repeat the part between the barriers from 10 to 50 times with a step of 10 to evaluate the fidelity change over time. 

\begin{figure}[h]
	\centering
	\begin{subfigure}{0.45\columnwidth}
		\centering
		\includegraphics[scale=0.3]{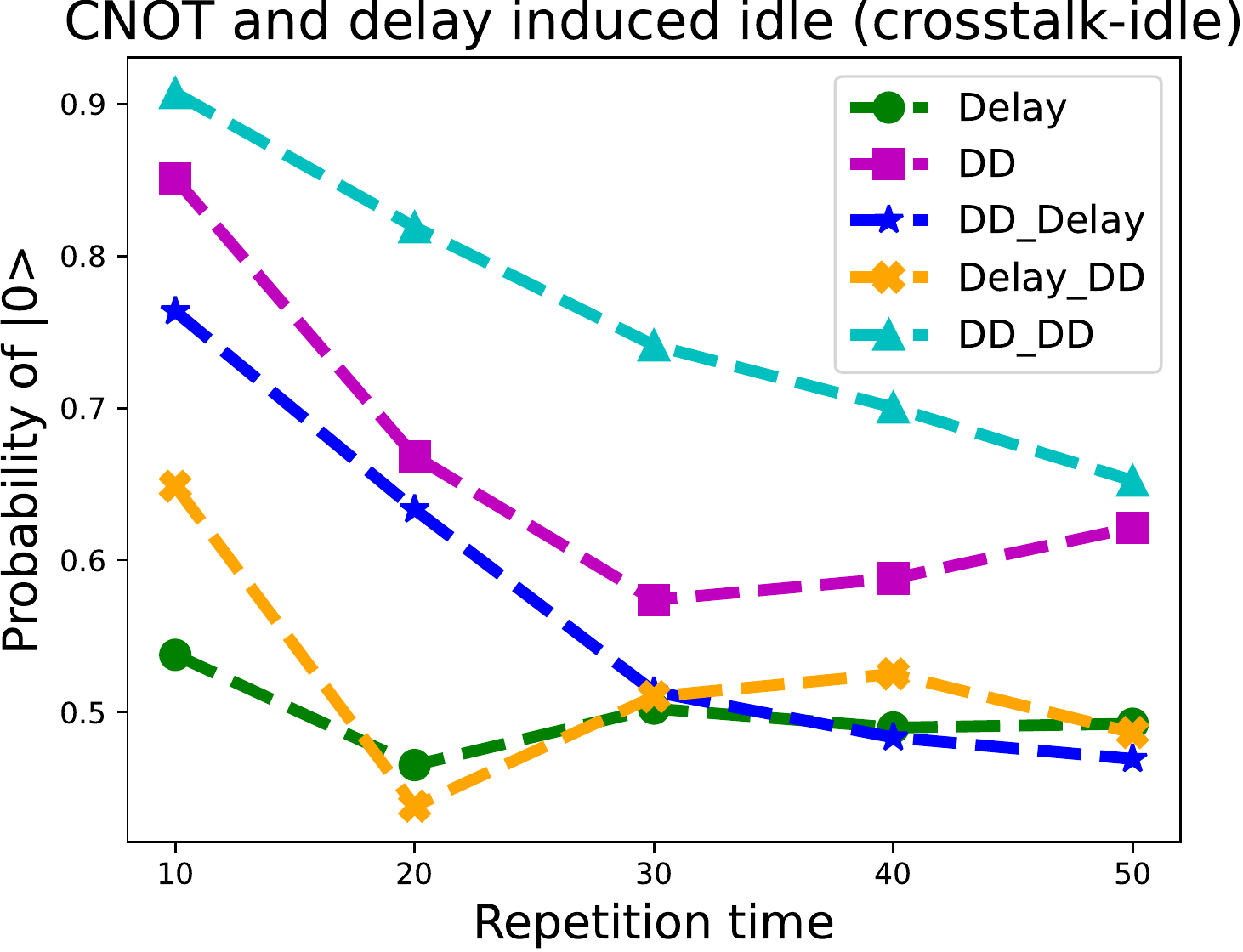}
		\label{fig:DD-cnot-delay-crosstalk}
		\caption{}
	\end{subfigure}
	\hfil
	\begin{subfigure}{0.45\columnwidth}
		\centering
		\includegraphics[scale=0.3]{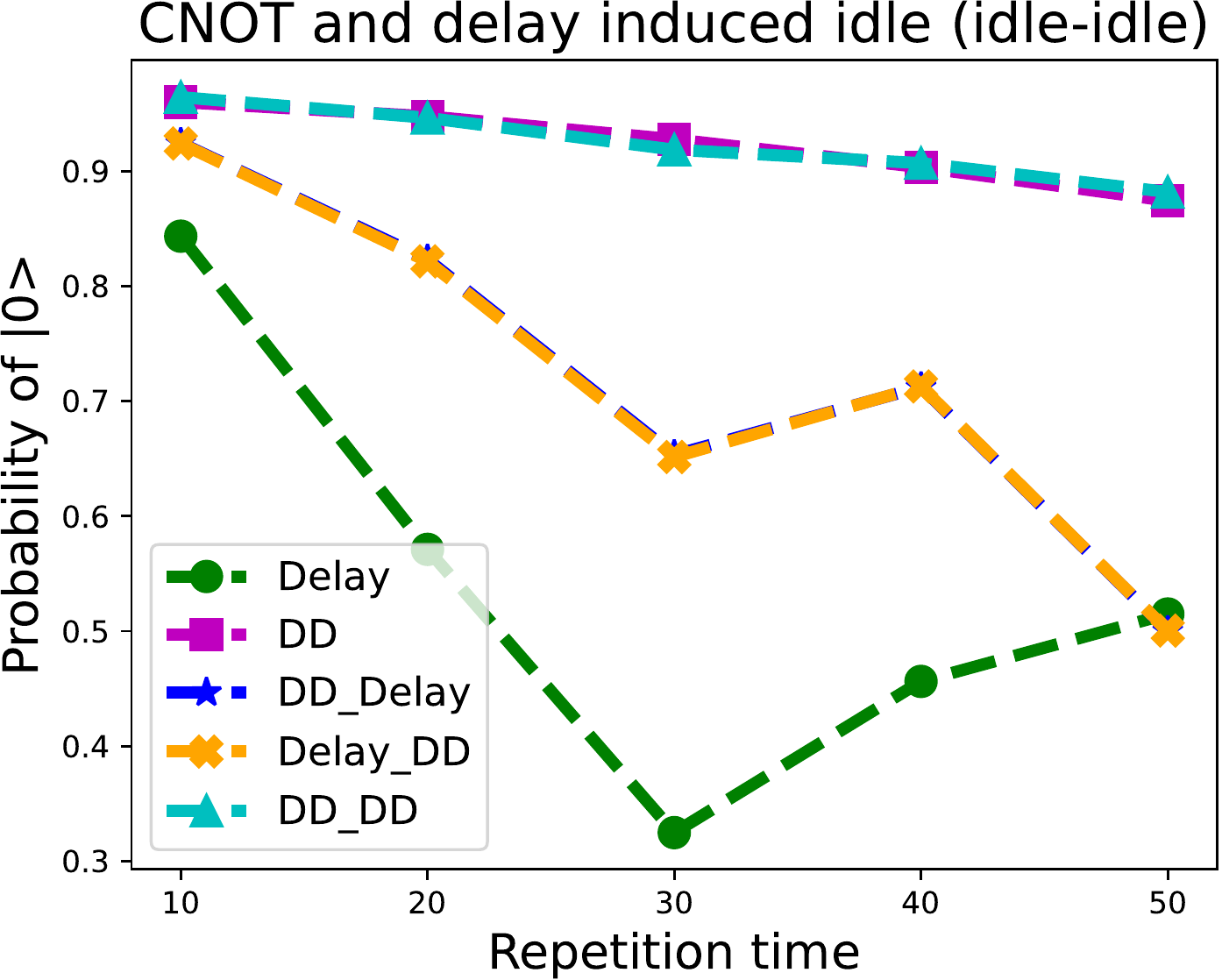}
		\label{fig:DD-cnot-delay-no-crosstalk}
		\caption{}
	\end{subfigure}
	
	\caption{Results of \texttt{CNOT} and other delay induced idle qubit when main qubit $q_0$ is considered as (a) crosstalk-idle qubit. (b) idle-idle qubit.
	}
	\label{fig:DD-cnot-delay}
	
\end{figure}

The results are shown in Fig.~\ref{fig:DD-cnot-delay}.
For crosstalk-idle qubit, double DD (DD\_DD) gives the best result and improves the fidelity by 15.7\% compared to single DD. DD\_Delay and Delay\_DD obtain unsatisfactory results. Delay\_DD has the worst results and is even sometimes worse than the baseline. Whereas for idle-idle qubit, double DD and single DD perform approximately the same and also have the best results. DD\_Delay and Delay\_DD obtain very close results and are better than the baseline.

\subsection{SWAP-induced Idle Qubit}
For \texttt{SWAP}-induced idle qubit, we mainly focus on two cases: (1) Purely \texttt{SWAP}-induced idle qubit. (2) \texttt{SWAP} and other delay induced idle qubit. Note that, since \texttt{SWAP} gate is composed of three \texttt{CNOT} gates, exploring the first case also indicates the DD insertion strategy for successive \texttt{CNOTs}. 

\begin{figure}[h]
	\centering
	\begin{subfigure}{0.45\columnwidth}
		\centering
		\includegraphics[scale=0.3]{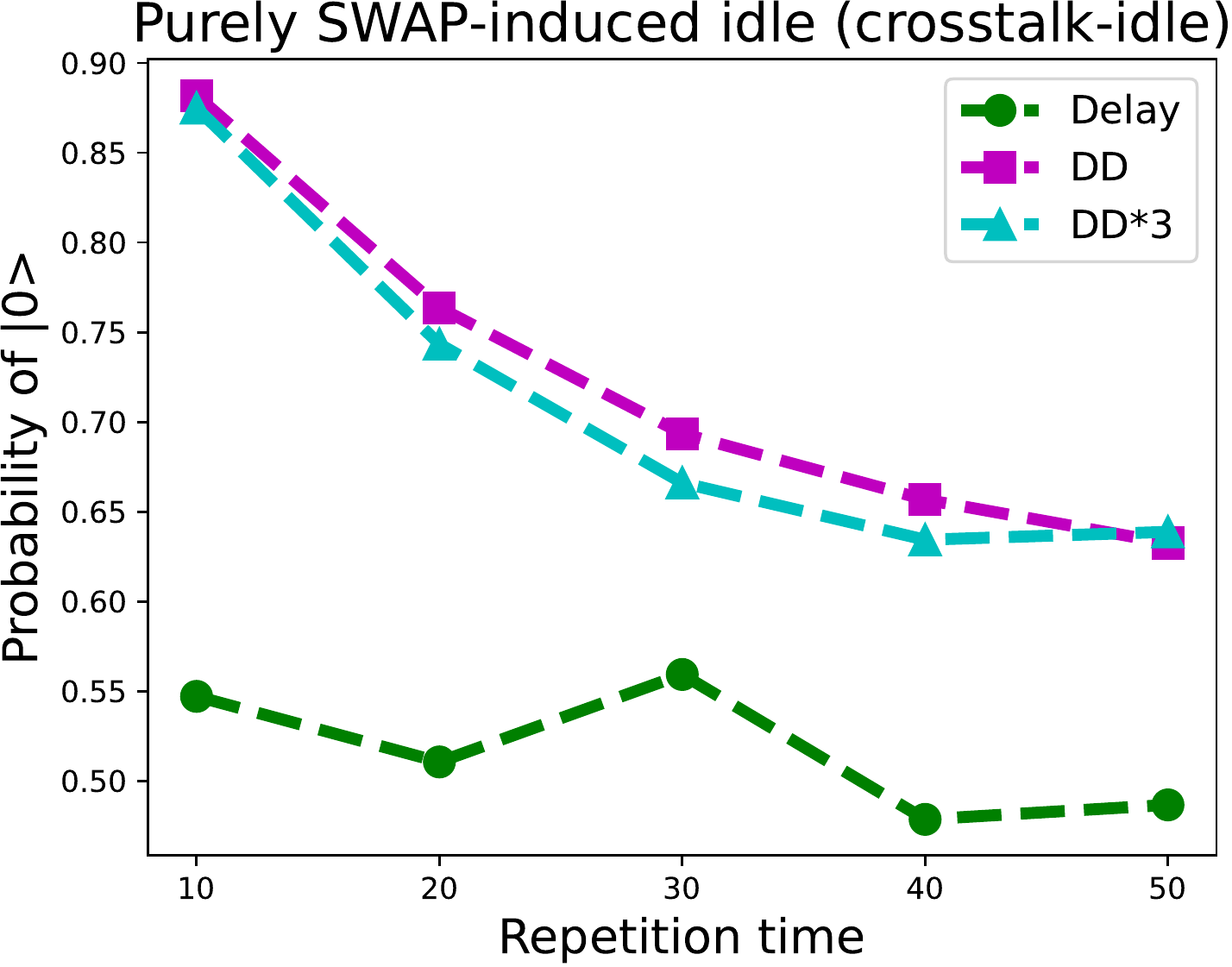}
		\label{fig:DD-swap-crosstalk}
		\caption{}
	\end{subfigure}
	\hfil
	\begin{subfigure}{0.45\columnwidth}
		\centering
		\includegraphics[scale=0.3]{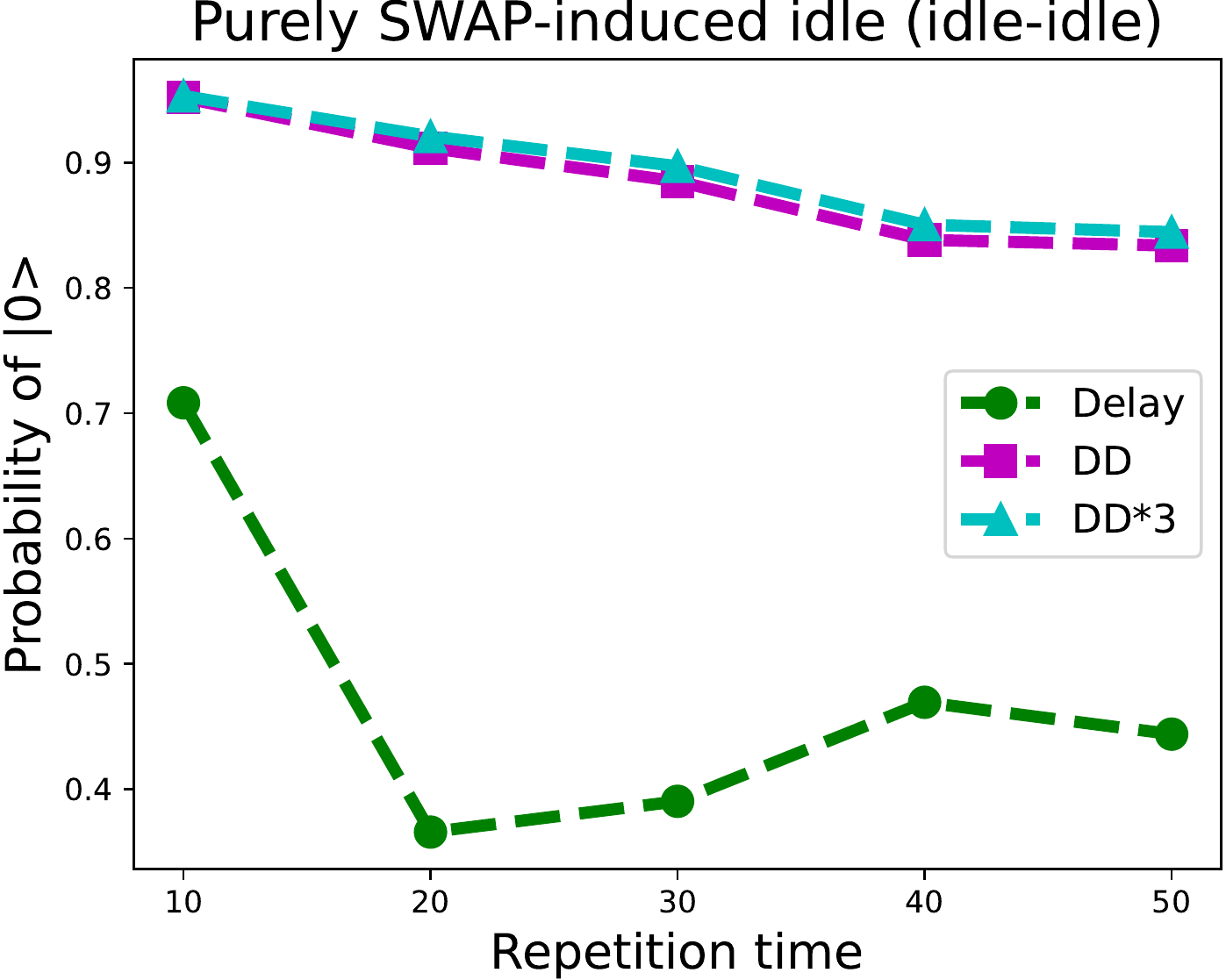}
		\label{fig:DD-swap-no-crosstalk}
		\caption{}
	\end{subfigure}
	
	\caption{Results of purely \texttt{SWAP} induced idle qubit when main qubit $q_0$ is considered as (a) crosstalk-idle qubit. (b) idle-idle qubit.
	}
	\label{fig:DD-swap}
	
\end{figure}

For purely \texttt{SWAP}-induced idle time, there are two strategies: (1) Insert a sequence of DD pulses for the overall idle time (labeled as DD). (2) Insert DD pulses for each \texttt{CNOT} of the \texttt{SWAP} gate independently (labeled as DD*3). In other words, three sequences of DD pulses are inserted. The results are shown in Fig.~\ref{fig:DD-swap}. For crosstalk-idle qubit, inserting DD only once during the \texttt{SWAP} duration performs slightly better than inserting it three times. Both cases outperform the baseline. Whereas for idle-idle qubit, DD*3 or single DD gives similar results. It shows that the increase of DD sequences cannot always bring benefits for the circuit fidelity.

\begin{figure}[h]
	\centering
	\begin{subfigure}{0.45\columnwidth}
		\centering
		\includegraphics[scale=0.3]{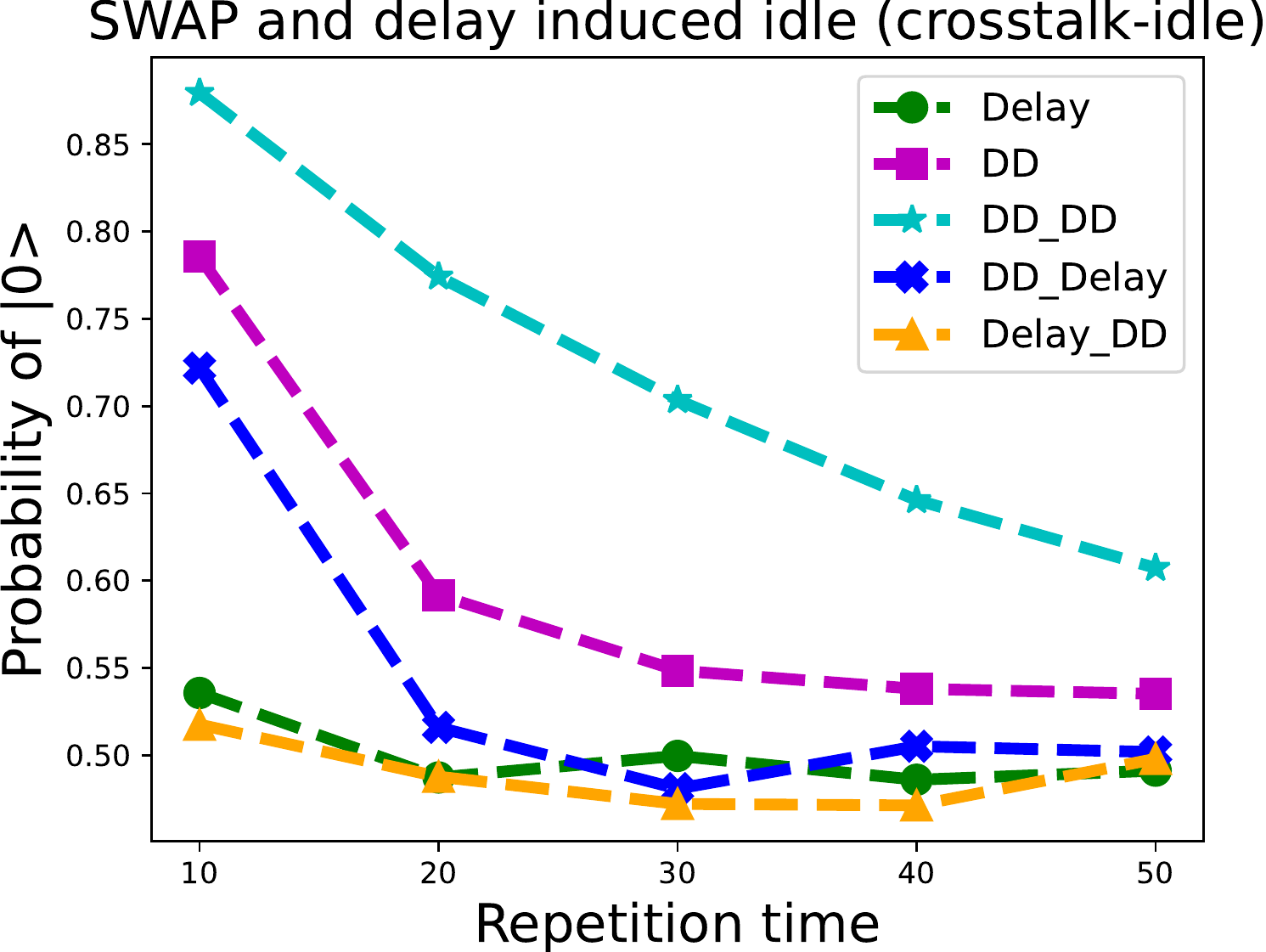}
		\label{fig:DD-SWAP-delay-crosstalk}
		\caption{}
	\end{subfigure}
	\hfil
	\begin{subfigure}{0.45\columnwidth}
		\centering
		\includegraphics[scale=0.3]{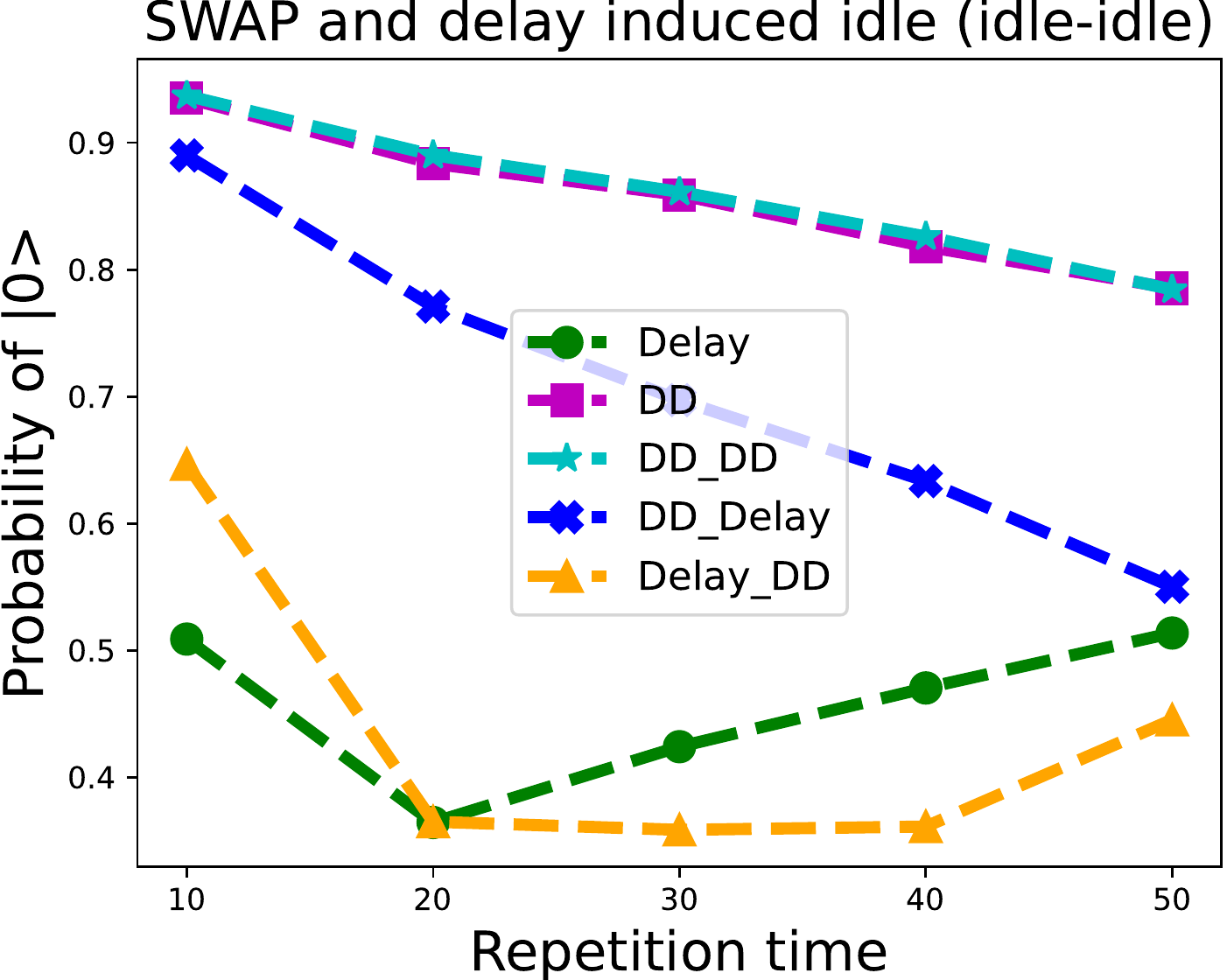}
		\label{fig:DD-SWAP-delay-no-crosstalk}
		\caption{}
	\end{subfigure}
	
	\caption{Results of \texttt{SWAP} and othe delay induced idle qubit when main qubit $q_0$ is considered as (a) crosstalk-idle qubit. (b) idle-idle qubit.
	}
	\label{fig:DD-SWAP-delay}
	
\end{figure}

While for \texttt{SWAP} and other delay induced idle qubit, there are also four DD insertion strategies, similar as \texttt{CNOT} and other delay induced idle qubit as explained in section~\ref{sec:cnot-idle} but replacing \texttt{CNOT} by \texttt{SWAP}.
Since inserting DD pulses for the total \texttt{SWAP} duration performs better than inserting three independent DD pulses based on the previous experimental results,  we only insert DD pulses once for \texttt{SWAP} gate. The results are shown in Fig.~\ref{fig:DD-SWAP-delay}. Similarly as the experiments for \texttt{CNOT} plus delay induced idle qubit, double DD performs the best for crosstalk-idle qubit and the fidelity is improved by 20.4\% compared with single DD. While both single DD and double DD achieve the best performance for idle-idle qubit. Delay\_DD keeps being the worst case for idle-idle and crosstalk-idle qubit.

\section{Ramsey experiment}

\begin{figure}[!h]
	\centering
	\begin{subfigure}{0.45\columnwidth}
		\centering
		\includegraphics[scale=0.4]{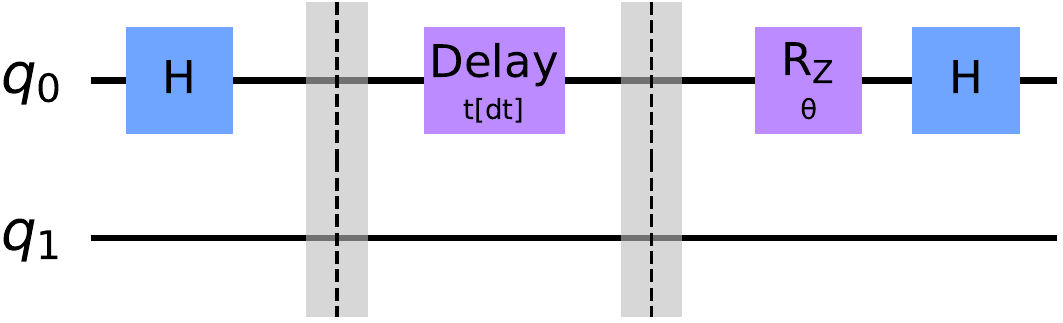}	
		\caption{}
	\end{subfigure}
	\hfill
	\begin{subfigure}{0.45\columnwidth}
		\centering
		\includegraphics[scale=0.4]{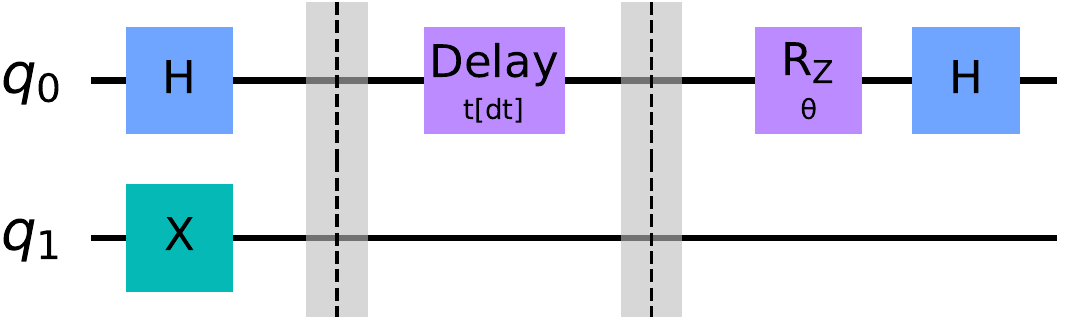}
		\caption{}
	\end{subfigure}
	
	\begin{subfigure}{0.45\columnwidth}
	\centering
	\includegraphics[scale=0.4]{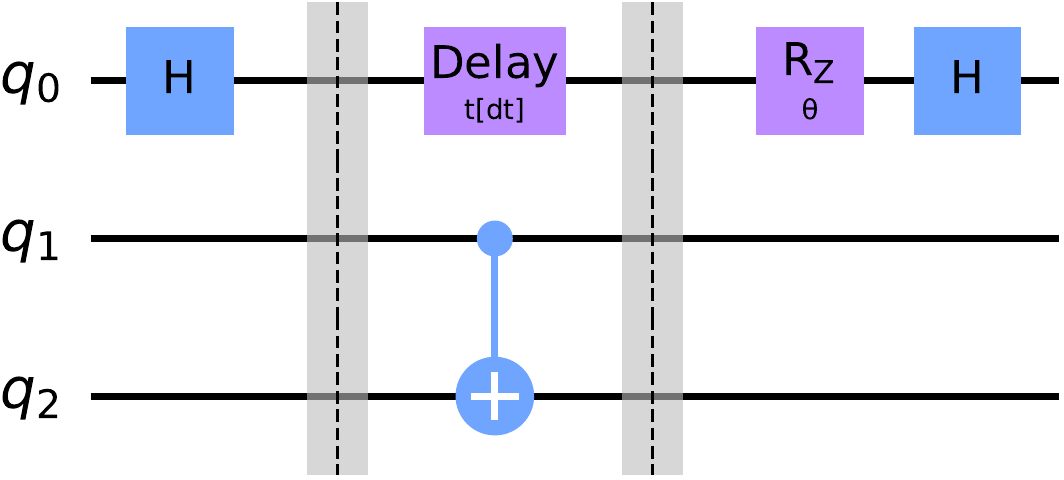}
	\caption{}
	\end{subfigure}
	\hfill
	\begin{subfigure}{0.45\columnwidth}
		\centering
		\includegraphics[scale=0.4]{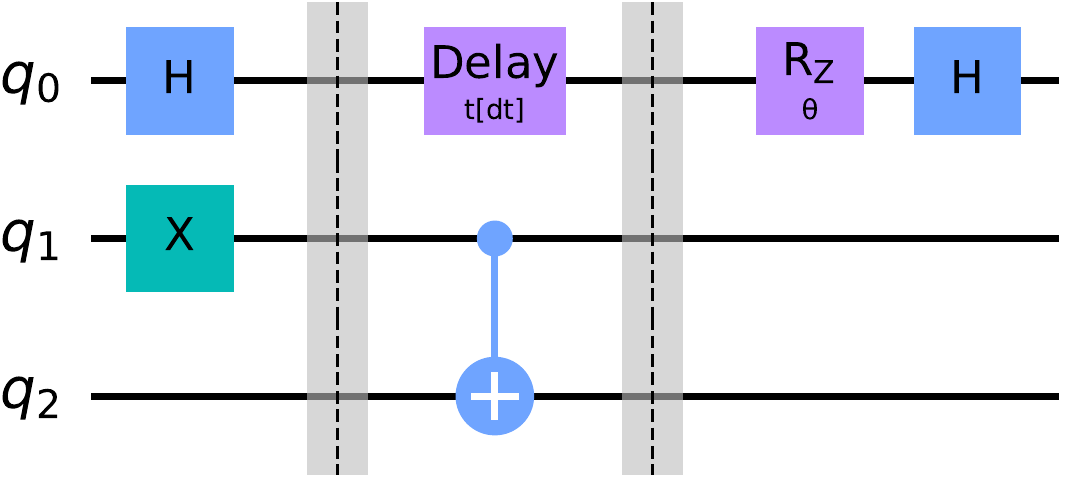}
		\caption{}
	\end{subfigure}
	\caption{Ramsey experiment circuits.
	}
	\label{fig:ramsey}
	
\end{figure}

In order to explore the reasons why double DD performs better than single DD for crosstalk-idle qubit, we characterize the crosstalk impact on idle qubit by performing Ramsey experiments to evaluate the frequency changes.

There are two types of crosstalk. The first one is an always-on-ZZ interaction between coupled qubits~\cite{kandala2021demonstration}. The second is due to the simultaneous operations and the most significant crosstalk is caused by the two-qubit cross-resonance gate~\cite{magesan2020effective}. We perform four types of Ramsey experiments to measure the detuning frequency of the main qubit: (1) a two-qubit circuit with $q_1$ in $|0\rangle$ (Fig.~\ref{fig:ramsey}(a)). (2) a two-qubit circuit with $q_1$ in $|1\rangle$ (Fig.~\ref{fig:ramsey}(b)). (3) a three-qubit circuit where $q_1$ is initialized in $|0\rangle$ and \texttt{CNOT} gate is applied between $q_1$ and $q_2$ (Fig.~\ref{fig:ramsey}(c)). (4) a three-qubit circuit where $q_1$ is initialized in $|1\rangle$ and \texttt{CNOT} gate is applied between $q_1$ and $q_2$ (Fig.~\ref{fig:ramsey}(d)). The main qubit $q_0$ is first set to a superposition state and then performs a free evolution marked as a \emph{Delay}.
 The duration of the \emph{Delay} is set to match the simultaneous \texttt{CNOT} duration. 
 After the delay, the main qubit adds a phase gate to better measure the detuning frequency and another $H$ gate in the end. We construct a list of circuits consisting of 1 to 50 repetitions of the middle part of the circuit inside the barriers to evaluate the main qubit decoherence across time.

\begin{figure}
	\centering
	\begin{subfigure}{0.45\columnwidth}
		\centering
		\includegraphics[scale=0.3]{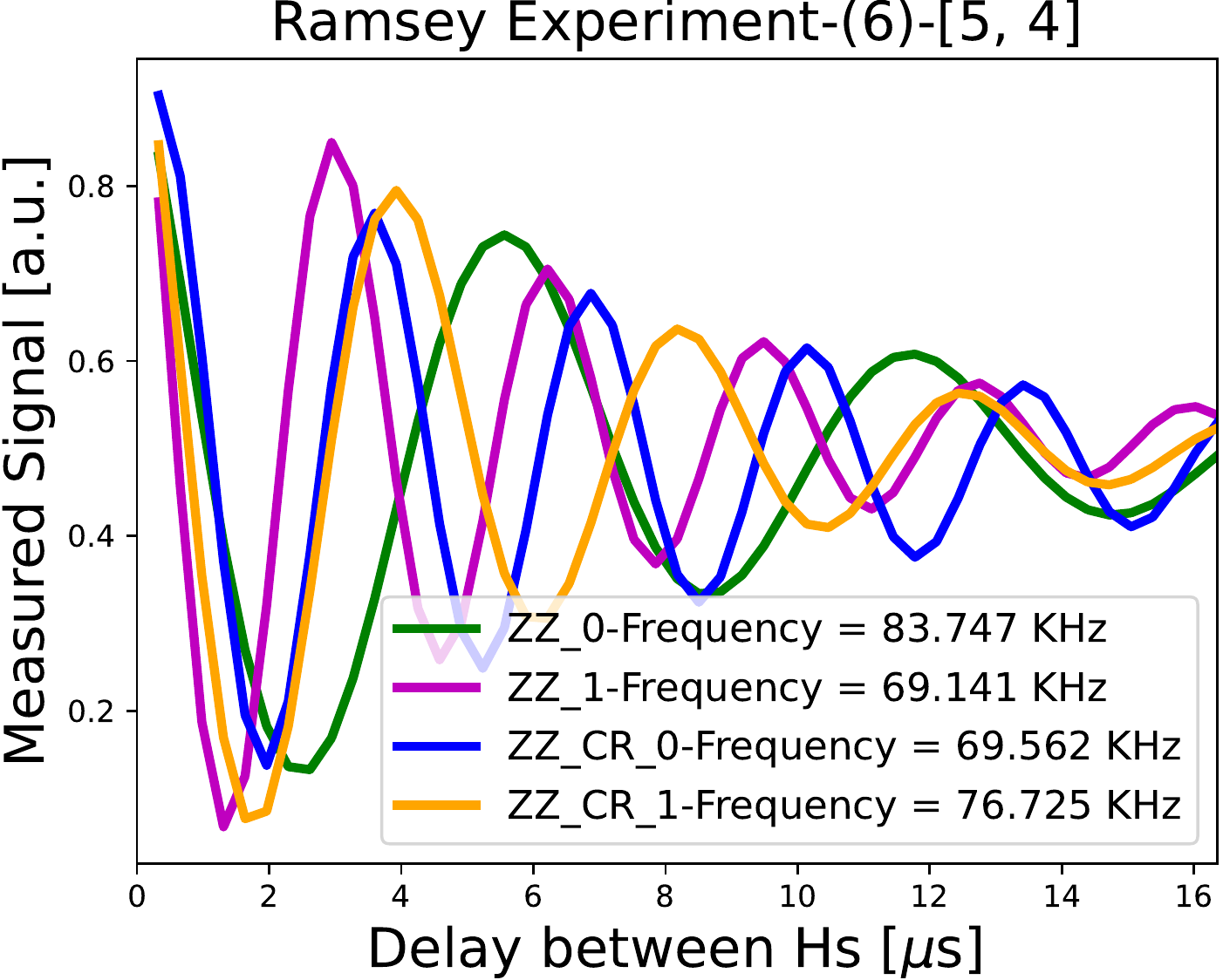}
		\label{fig:id1}
		\caption{}
	\end{subfigure}
	\hfil
	\begin{subfigure}{0.45\columnwidth}
		\centering
		\includegraphics[scale=0.3]{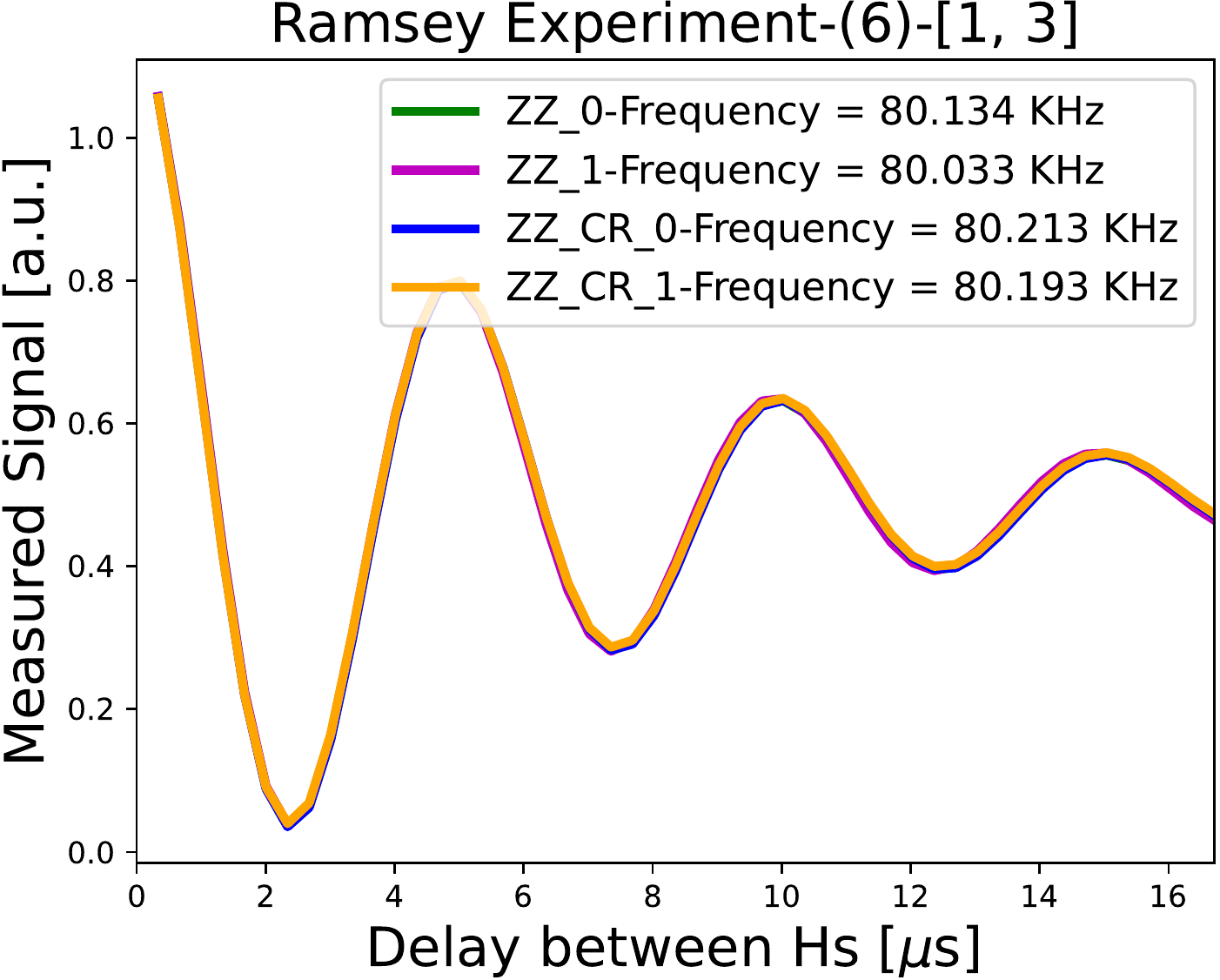}
		\label{fig:id2}
		\caption{}
	\end{subfigure}
	
	\caption{Results of Ramsey experiment by mapping the main qubit $q_0$ to $Q_6$ and mapping $q_1$, $q_2$ to (a) $Q_5$, $Q_4$. (b) $Q_1$, $Q_3$.
	}
	\label{fig:ramsey_result}
	
\end{figure}

We show the results of mapping the main qubit $q_0$ to $Q_6$ on IBM Q 7 Lagos. We first map $q_1$ and $q_2$ to $Q_5$ and $Q_4$ on the hardware to physically couple $q_0$ and $q_1$. We can see a frequency difference of around 14.6 kHz between the first two cases, which is due to the always-on-ZZ interaction. A frequency difference of 14.2 kHz exists between the first and the third case, which illustrates the impact of cross-resonance induced crosstalk on changing the adjacent qubit frequency. Whereas when we decouple the qubits by mapping $q_1$ and $q_2$ to $Q_1$ and $Q_3$, in other words, no qubit is physically connected with the main qubit in the circuit, the frequency of the main qubit keeps almost the same as shown in Fig.~\ref{fig:ramsey_result}(b). We have results with the same trend by mapping the main qubit to each physical qubit and the spectator qubits to all the connected qubits. Note that, for each experiment, the four types of circuits are executed in one job so that the calibration data keep unchanged. 

Since DD is composed of $\pi$ pulses, if there is a frequency shift on the DD-target qubit, it might introduce deviations in $\pi$ rotations such that the idle qubit cannot come back to its original state after DD insertion, which explains the reason why double DD performs better than single DD for crosstalk-idle qubit.

\section{Discussion}
As idle qubits frequently appear in superconducting circuits, as well as the severe impact of decoherence error, especially for large circuits, it is important to assess the optimal strategy to insert DD sequences under different conditions to mitigate the decoherence error. Here, we provide some guidelines that can help the community better insert DD pulses to build circuits with high fidelity:

\begin{itemize}
	\item For crosstalk-idle qubit, if it is induced by \texttt{CNOT} or \texttt{SWAP} along with some other delay, it is recommended to insert DD sequences during \texttt{CNOT}/\texttt{SWAP} and the other delay separately. 
	
	\item If the crosstalk-idle qubit is induced by a sequence of \texttt{CNOT} or \texttt{SWAP}, only applying DD once during the overall idle time can achieve good results.

	\item For idle-idle qubits, inserting DD only once during the total idle time is enough to improve the circuit fidelity significantly.
	 
\end{itemize}

In our paper, we use small circuits to understand the impact of different DD insertion strategies. Based on our observations and guidelines, our next step is to explore the DD insertion strategies for large circuits and build a general DD-insertion compiler to mitigate decoherence error without the overhead of executing additional circuits. Also, we only explore CPMG pulses due to its simplicity and single-axis $\pi$ rotation. It is interesting to analyze other more complex DD techniques with extra phase shift such as $XY4$. These techniques can further contribute to the pulse-level circuit optimization~\cite{shi2019optimized,liang2022variational}.

\section{Conclusion}
Today's quantum computer performs error-prone quantum operations with decoherence error being one of the primary error sources. In this paper, we focus on the decoherence error mitigation and explore the generally used method -- dynamical decoupling. The idle qubit is caused under different conditions and sometimes influenced by crosstalk depending on the qubit coupling. First, we analyze \texttt{CNOT} and \texttt{SWAP} induced idle qubit and further divide it into crosstalk-idle and idle-idle. Second, we examine DD insertion strategies for the two types of idle qubits under the analyzed configurations. Finally, we provide guildelines of DD insertion strategies and help the community design DD-protected circuits to achieve reliable results. Our work can be regarded as the first step to developing a general DD-insertion compiler for decoherence error mitigation. 

\section*{Acknowledgment}
This work is funded by the QuantUM Initiative of the
Region Occitanie, University of Montpellier and IBM
Montpellier. The authors sincerely appreciate the discussions with Dr. Nick Bronn and are very grateful to Adrien Suau for the helpful suggestions.

\bibliographystyle{plain}

\bibliography{bibliography}{}

\end{document}